# A probability density for modeling unknown physical processes

Steven C. Gustafson[1] and Adam C. Hillier[2]
Paper submitted 19 April 2011 to arXiv.org (general physics)
Air Force Institute of Technology, Ohio, USA
[1]retired faculty, [2]graduate student

**Abstract**

This brief paper develops a probability density that models processes for which the physical mechanism is unknown. It has desirable properties which are not realized by densities derived from Gaussian process or other classic methods. In many areas of physics, such the analysis of chaotic phenomena, there is need for a density that has these properties.

**Properties of the density**

Consider a process that generates an output y for any input x, where the generation mechanism is unknown. Points $(x_i, y_i)$, i = 1, 2, …, n are available which give the output generated for some inputs. These points are to be used to predict the output that would have been generated for other inputs. In particular, predictions are to be made using a probability density of y given x which has the following properties, where classic definitions, e.g., for roughness, for point centered Gaussian kernels, and for the quadratic variance of the least squares line, are employed as discussed in [1, 2].

1. The density is a delta function at the $x_i$. Thus if x is at a point input, then y is the point output.

2. The density extrapolates to a mean versus x which is the least squares line of the points and to a variance versus x which is the classic quadratic function associated with this line. Thus a simple linear model applies at large distance from the points.

3. The density has a mode function which is the least squares line plus a classic sum of point-centered Gaussian kernels of the same variance $w^2$, where this variance is such that the function has least classic roughness. Thus the trace of the probability density maximum has a simple form and has maximum smoothness.

4. The density has a skew normal form with mean and variance functions which are the mean and variance of two basis functions, where these functions are the plus and minus square root of the classic quadratic variance for the least squares line plus a sum of Gaussian kernels. The kernels have the same variance for each function and are $w_1^2$ and $w_2^2$. Thus the density (1) is in general skewed, (2) is determined by the smallest possible number of simple basis functions, and (3) is characterized by parameters $w_1$, and $w_2$.

**Specification of the density**

To simplify the specification without loss of generality, let the points be normalized so that they have the x axis as their least squares line and so that they have zero mean and unit variance in both x and y (the normalization can be inverted to apply the specification to the original points).

Then the mode function is $m(x) = \sum A_i \exp[-(x - x_i)^2/(2w^2)]$, where the $A_i$ are such that the points are intersected, i.e., $m(x_i) = y_i$, and w is such that roughness is least, i.e., $\int_{-\infty}^{\infty} m''^2(x)\,dx$ is minimized.

Also, the two basis functions are $f(x) = (1 + 1/n + x^2/n)^{1/2} + \sum B_i \exp[-(x - x_i)^2/(2w_1^2)]$ and $g(x) = -(1 + 1/n + x^2/n)^{1/2} + \sum C_i \exp[-(x - x_i)^2/(2w_2^2)]$, where the $B_i$ and $C_i$ are such that the points are intersected, i.e., $f(x_i) = y_i$ and $g(x_i) = y_i$, and where $1 + 1/n + x^2/n$ is the classic quadratic variance for the least squares line.

It is easily verified that the mean function $\mu(x) = [f(x) + g(x)]/2$ intersects the points and extrapolates to zero and that the variance function $\sigma^2(x) = \{[f(x) - \mu(x)]^2 + [g(x) - \mu(x)]^2\}/2$ is zero at the points and extrapolates to $1 + 1/n + x^2/n$, which is the classic variance associated with the least squares line.

Figures 1 and 2 show the mode function between two basis functions for three points which are normalized as described above: $(-(3/2)^{1/2}, -(1/2)^{1/2})$, $(0, 2(1/2)^{1/2})$, and $((3/2)^{1/2}, -(1/2)^{1/2})$. Here $w = 1.391$, $w_1 = 1.936$, and $w_2 = 1.664$, where (1) w is such that the roughness of the mode function is minimized, as required, (2) $w_1$ is chosen to be the largest value such that $\sigma^2(x)$ increases monotonically beyond the data end points, and (3) $w_2$ is chosen to be the mean of w and $w_1$. Note that the latter two choices are for the purpose of illustration and that alternative choices could be made; e.g., $w_1$ and $w_2$ could be set equal to w. If $w_1$ does not equal $w_2$, then as shown in Figure 2 the mean function (which is the mean of the two basis functions) does not equal the mode function (except at the points or at large magnitude x), and thus the density is skewed.

The probability density is then the skew normal density of y given x determined by the mode, mean, and variance functions. This density [3] is $p(y|x) = (2\pi)^{-1/2}(1/w)\exp[-(y-a)^2/(2b^2)]\{1 + \text{erf}[c(y-a)/(\sqrt{2}b)]\}$, where $\mu = a + (2/\pi)^{1/2} bc(1+c^2)^{-1/2}$, $\sigma^2 = b^2[1 - (2/\pi)c^2/(1+c^2)]$, and here $\mu$, $\sigma^2$, and the parameters a, b, and c are functions of x. Taking the derivative of $p(y|x)$ with respect to y and setting it to zero and letting m be the y that satisfies the resulting equation yields
$[(m - a)/b^2]\{1 + \text{erf}[c(m-a)/(\sqrt{2}b)]\} = (2/\pi)^{1/2}(c/b)\exp[-c^2(m-a)^2/(2c^2)]$, where m is a function of x. There are thus three nonlinear equations that determine a, b, and c and therefore $p(y|x)$ in terms of m, $\mu$, and $\sigma^2$.

### Remarks on the density

Various standard methods, and in particular Gaussian process [4] methods, can specify probability densities that realize some of the properties listed above. However, no standard method realizes a density that is in general skewed, and a skewed density is clearly more realistic than a symmetric density, especially in regions where the mode function has large curvature. In particular, any Gaussian process method must specify a Gaussian density of y given x, and this density may be interpreted as the resulting from an infinite number of basis functions. In contrast, the density described here results from only a mode function plus two basis functions.

The density described here is not computationally intensive. The mode function requires the solution of n linear equations in n unknowns to determine the $A_i$ for each of a number of candidate values of w until the value that minimizes roughness is found. For a given $w_1$ or $w_2$, the corresponding basis function requires only the solution of n linear equations in n unknowns, and for a given x finding the skew normal density requires only the solution of three nonlinear equations in three unknowns.

The density described here may be important in that prediction using a density that has its properties is a need encountered in many areas of physics, such the analysis of chaotic phenomena [5].

**References**

1. S. C. Gustafson, D. R. Parker, and R. K. Martin, "Cardinal interpolation", *IEEE Transactions on Pattern Analysis and Machine Intelligence,* vol. 29, pp. 1538 -1545, 2007.

2. M. R. Falknor, E. M. Guild, A. C. Hillier, E. C. Like, and S. C. Gustafson, "Direct cardinal interpolation", *Electronic Journal of Applied Statistical Analysis*, vol. 3, pp. 123-131, 2010.

3. A. Azzalini, "A class of distributions which includes the normal ones", *Scand. J. Statist.,* vol. 12, pp. 171-178, 1985.

4. C. Rasmussen and C. K. I. Williams, *Gaussian Processes for Machine Learning*, MIT Press, 2006.

5. R. L. Devaney, *Introduction to Chaotic Dynamical Systems*, Westview Press, 2003.

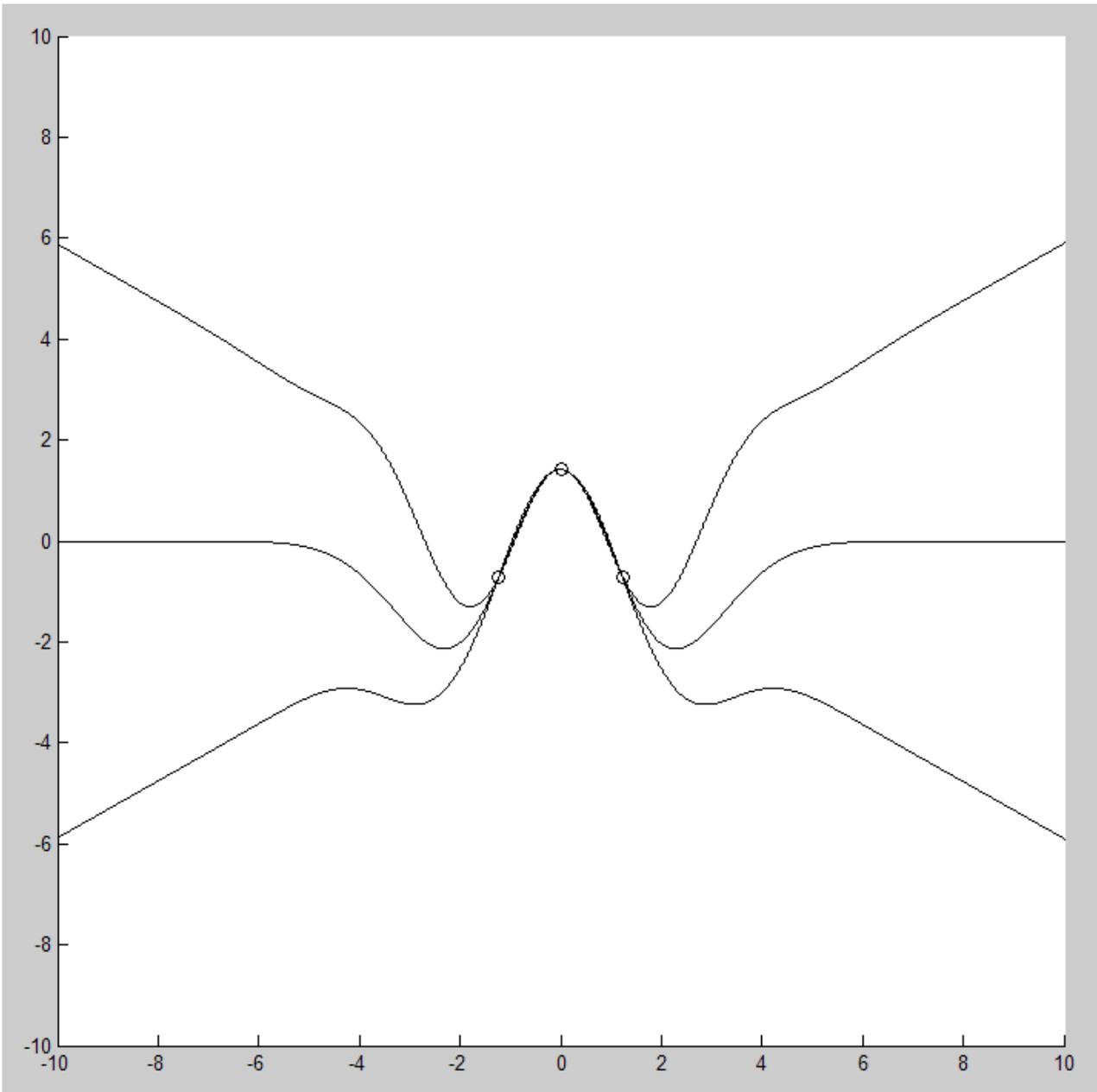

Figure 1. The mode function between two basis functions for three normalized points. The three points are $(-(3/2)^{1/2}, -(1/2)^{1/2})$, $(0, 2(1/2)^{1/2})$, and $((3/2)^{1/2}, -(1/2)^{1/2})$ and are normalized to have the x axis as their least squares line and to have zero mean and unit variance in x and y. The mode function is the sum of Gaussian kernels which intersects the points and which has least roughness, where the kernel standard deviation that minimizes roughness is $w = 1.391$. Each basis function also intersects the points and is a sum of Gaussian kernels plus or minus the square root of the classic quadratic variance for the least squares line. The standard deviation for the upper basis function is chosen to be the largest value such that the variance function increases monotonically beyond the data end points and is $w_1 = 1.936$. The standard deviation for the lower basis function is chosen to be the mean of w and $w_1$ and is $w_2 = 1.664$. Note that the latter two choices are for the purpose of illustration and that alternative choices could be made; e.g., $w_1$ and $w_2$ could be set equal to w. If $w_1$ does not equal $w_2$, then the mean function (which is the mean of the two basis functions) does not equal the mode function (except at the points or at large magnitude x), and thus the density is skewed.

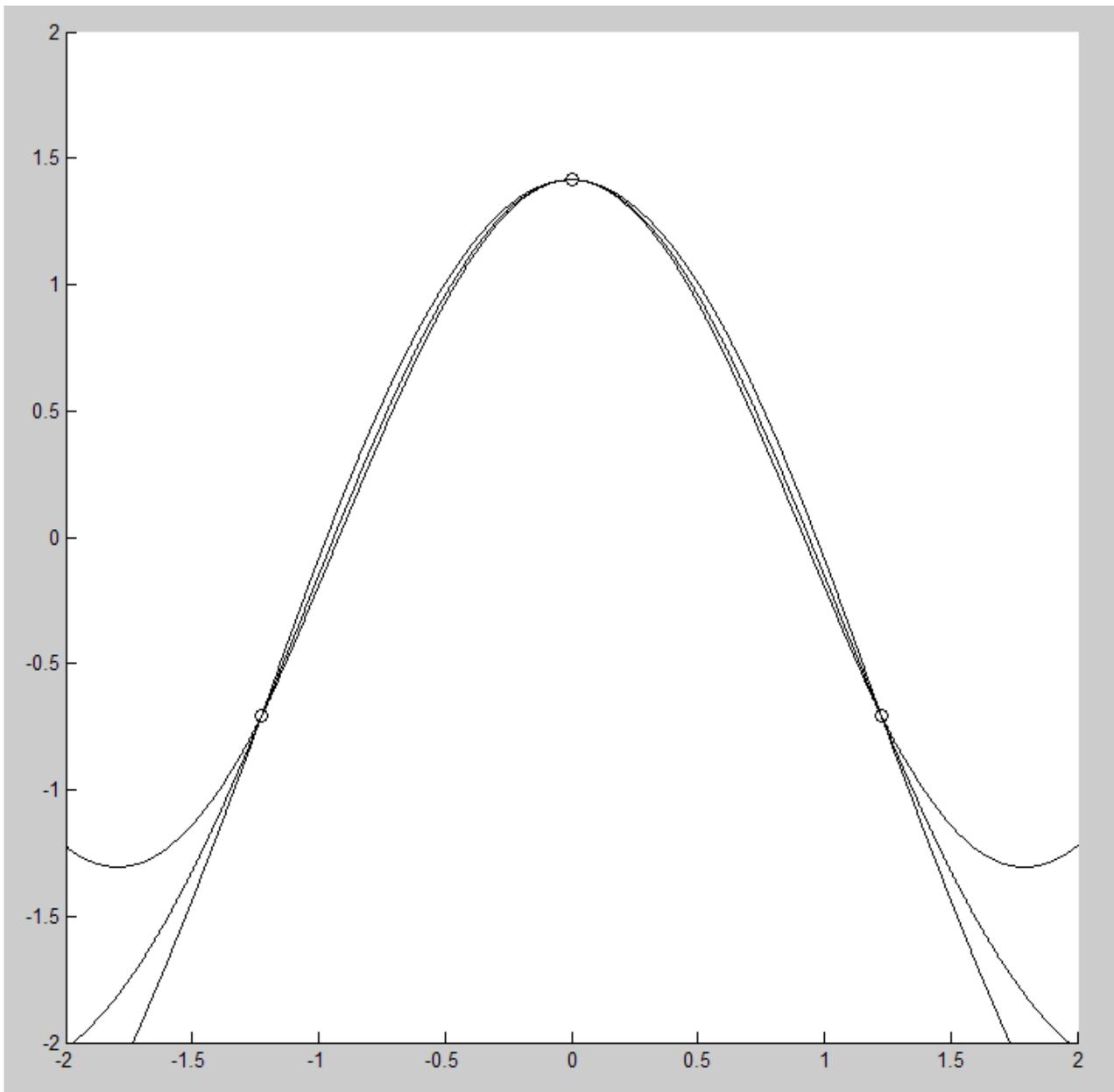

Figure 2. A magnification of Figure 1 which shows that the density is skewed, i. e, the mode function (which is between the two basis functions) clearly does not equal the mean function (which is the mean of the two basis functions).